\documentclass[letter]{ptptex}

\usepackage{graphicx}

\title{
Magnetic Properties of Precession Modes Built on High-$K$ Multi-quasiparticle States 
in $^{178}$W%
}

\author{
Masayuki \textsc{Matsuzaki}$^{1,}$\footnote{E-mail: matsuza@fukuoka-edu.ac.jp}%
and
Yoshifumi R. \textsc{Shimizu}$^{2,}$\footnote{E-mail: yrsh2scp@mbox.nc.kyushu-u.ac.jp}%
}

\inst{
$^1$Department of Physics, Fukuoka University of Education, \\
Munakata 811-4192, Japan
\\
$^2$Department of Physics, Graduate School of Sciences, Kyushu University, \\
Fukuoka 812-8581, Japan
}

\abst{
 We present an example that shows that the random phase approximation performed 
on high-$K$ 
multi-quasiparticle configurations leads to a rotor picture by calculating 
excitation energies and magnetic properties of $^{178}$W. Then we deduce the 
effective $g_R$ of the high-$K$ rotors and compare it with that of the low-$K$ 
one.
}

\begin{document}

\maketitle

 Rotation is one of typical collective motions in atomic nuclei. Normally axially 
symmetric nuclei carry large angular momenta in the form of collective rotation 
about an axis perpendicular to the symmetry axis. In some cases in which single 
particle orbitals with large angular momenta $j_i$ and their projections to the 
symmetry axis $\Omega_i$ sit at the vicinity of the Fermi surface --- realized 
typically in the $A \sim 180$ region ---, the nucleus can have large angular 
momenta by aligning multiple quasiparticles (QPs) along the symmetry axis. 
Sometimes the latter scheme forms yrast states, or even if not, isomers are often 
formed thanks to largeness of $K=\sum_i\Omega_i$.

 Detailed information about high-$K$ configurations can be obtained from their 
magnetic properties --- static magnetic moments and/or 
$g$-factors inferred from $B(M1)/B(E2)$ branching ratios of in-band 
transitions in rotational bands excited on top of high-$K$ configurations. 
The latter data are transformed into $|g_K-g_R|/Q_0$ by 
way of the rotor model~\cite{bm}. Then by assuming appropriate $g_R$ and $Q_0$, 
the extracted $g_K$ is compared with the weighted average of single-$j$ $g$-factors 
with respect to $\Omega_i$.~\cite{re177} 

 On the other hand, (at least lower members of) rotational bands excited on 
high-$K$ configurations can be described as multiple excitations of the 
precession phonons in the language of the random phase approximation 
(RPA)~\cite{ander,skal}. Thus, by calculating the wave function of the one phonon 
state, $B(M1: I=K+1\rightarrow K)$ can be obtained and transformed into the 
effective (RPA) $(g_K-g_R)$; its sign can be determined by the calculated 
$E2/M1$ mixing ratio. The magnetic moment $\langle \mu \rangle$ and 
accordingly the $g$-factor, 
$g=\sqrt{\frac{4\pi}{3}}\langle \mu \rangle/(\langle J \rangle\mu_N)$, 
of high-$K$ configurations can be calculated in the mean field level. 
Since this $g$ essentially coincides with $g_K$, we can deduce $g_R$ 
of the high-$K$ rotor by combining the RPA $(g_K-g_R)$ and $g$. 

 The purpose of the paper is twofold: By applying the above method to $^{178}$W 
for which the richest experimental information~\cite{w1,w2,w3} is available, 
we corroborate that the RPA gives a rotor picture via the excitation energies 
and magnetic properties, and deduce $g_R$ of high-$K$ rotors. 

 We begin with a one-body Hamiltonian, 
\begin{gather}
h'=h-\hbar\omega_\mathrm{rot}J_x, \notag \\
h=h_\mathrm{Nil}-\mit\Delta_\tau (P_\tau^\dagger+P_\tau)
                   -\lambda_\tau N_\tau , \notag \\
h_\mathrm{Nil}=\frac{\mathbf{p}^2}{2M}
                +\frac{1}{2}M(\omega_x^2 x^2 + \omega_y^2 y^2 + \omega_z^2 z^2)
                +v_{ls} \mathbf{l\cdot s}
                +v_{ll} (\mathbf{l}^2 - \langle\mathbf{l}^2\rangle_{N_\mathrm{osc}}) .
                \label{hnil}
\end{gather}
Here $\tau = 1$ and 2 stand for neutron and proton, respectively, 
and chemical potentials $\lambda_\tau$ are determined so as to give correct average 
particle numbers $\langle N_\tau \rangle$. The oscillator frequencies 
are related to the quadrupole deformation parameters $\epsilon_2$ and $\gamma$ 
in the usual way. [We adopt the so-called Lund convention.] 
The orbital angular momentum $\mathbf{l}$ is defined in the 
singly-stretched coordinates $x_k' = \sqrt{\frac{\omega_k}{\omega_0}}x_k$, 
with $k =$ 1 -- 3 denoting $x$ -- $z$, and the corresponding momenta. 
Nuclear states with QP excitations, \textit{i.e.,} alignments along 
the $x$ axis, are obtained by exchanging the QP energy and wave functions such as 
\begin{equation}
(-e'_\mu, \mathbf{V}_\mu, \mathbf{U}_\mu) \rightarrow 
(e'_{\bar\mu}, \mathbf{U}_{\bar\mu}, \mathbf{V}_{\bar\mu}) ,
\label{exch}
\end{equation}
where ${\bar\mu}$ denotes the signature partner of $\mu$. 

We perform the RPA to the residual pairing plus 
doubly-stretched quadrupole-quadrupole ($Q'' \cdot Q''$) interaction between QPs. 
Since we are interested in the precession mode that has a definite signature quantum number, 
$\alpha = 1$, only two components out of five of the $Q'' \cdot Q''$ interaction 
are relevant. They are $K = \pm1$ components. 
Note that we call the symmetry axis, with respect to which the $K$ quantum number 
is defined, the $x$ axis throughout this paper. That is, we consider the $\gamma=-120^\circ$ 
case.  These components of the interaction are related to the restoration
of the spherical symmetry.  Requiring the decoupling of this symmetry mode
(the Nambu-Goldstone mode), $J_\pm=J_y \pm iJ_z$,
the strength of the interaction is determined.
Then, utilizing the identities in Table III of Ref.~\citen{smptp}\footnote{
Strictly speaking, the identities have tiny deviations
brought about by the singly-stretched (rather than the non-stretched) 
$\mathbf{l\cdot s}$ and $\mathbf{l}^2$ potentials.}, 
the RPA equation of motion can be cast into~\cite{ma} the following form~(\ref{matrix}),
which we use in the actual calculation rather than the original
equation in terms of $Q''$ operators:
\begin{equation}
(\omega^2-\omega_\mathrm{rot}^2)\left|
\begin{array}{@{\,}cc@{\,}}
A(\omega) & C(\omega) \\
B(\omega) & D(\omega)
\end{array}\right|=0 ,
\label{matrix}
\end{equation}
where
\begin{gather}
A(\omega)=\omega\mathcal{J}_y(\omega)-\omega_\mathrm{rot}\mathcal{J}_{yz}(\omega) , \notag \\
B(\omega)=\omega_\mathrm{rot}\left(\mathcal{J}_y(\omega)-\mathcal{J}_x\right)
          -\omega\mathcal{J}_{yz}(\omega) , \notag \\
C(\omega)=\omega_\mathrm{rot}\left(\mathcal{J}_z(\omega)-\mathcal{J}_x\right)
          -\omega\mathcal{J}_{yz}(\omega) , \notag \\
D(\omega)=\omega\mathcal{J}_z(\omega)-\omega_\mathrm{rot}\mathcal{J}_{yz}(\omega) , 
\label{abcd}
\end{gather}
with
\begin{gather}
\mathcal{J}_x=\hbar\langle J_x \rangle/\omega_\mathrm{rot} , \notag \\
\mathcal{J}_y(\omega)=\sum_{\mu<\nu}^{(\alpha=\pm1/2)}\frac{2E_{\mu\nu}\left(iJ_y(\mu\nu)\right)^2}
                      {E_{\mu\nu}^2-(\hbar\omega)^2} , \notag \\
\mathcal{J}_z(\omega)=\sum_{\mu<\nu}^{(\alpha=\pm1/2)}\frac{2E_{\mu\nu}\left(J_z(\mu\nu)\right)^2}
                      {E_{\mu\nu}^2-(\hbar\omega)^2} , \notag \\
\mathcal{J}_{yz}(\omega)=\sum_{\mu<\nu}^{(\alpha=\pm1/2)}\frac{2\hbar\omega iJ_y(\mu\nu)J_z(\mu\nu)}
                      {E_{\mu\nu}^2-(\hbar\omega)^2} .
\label{inertia}
\end{gather}
Here we adopt the convention that matrix elements of $J_y$ and $\mu_y$ (below) are pure imaginary. 
The non-spurious part of Eq.~(\ref{matrix}), 
$A(\omega)D(\omega)-B(\omega)C(\omega) = 0$, can be rewritten as
\begin{equation}
\left[\omega\mathcal{J}_+^{(\mathrm{eff})}(\omega)
      -\omega_\mathrm{rot}\left(\mathcal{J}_x-\mathcal{J}_+^{(\mathrm{eff})}(\omega)\right)\right]
\left[\omega\mathcal{J}_-^{(\mathrm{eff})}(\omega)
      +\omega_\mathrm{rot}\left(\mathcal{J}_x-\mathcal{J}_-^{(\mathrm{eff})}(\omega)\right)\right]
=0 ,
\label{disp_p}
\end{equation}
where the suffixes $+$ and $-$ refer to the $\Delta K = +1$ and $-1$ modes, respectively, and
\begin{gather}
\mathcal{J}_\pm^{(\mathrm{eff})}(\omega)=\mathcal{J}_\perp(\omega)\mp\mathcal{J}_{yz}(\omega) 
, \notag \\
\mathcal{J}_\perp(\omega)=\mathcal{J}_y(\omega)=\mathcal{J}_z(\omega) .
\label{effiner}
\end{gather}
For $\Delta K = +1$ excitations, corresponding to the precession modes,
the excitation energy in the laboratory frame is given by
\begin{equation}
\hbar\omega+\hbar\omega_\mathrm{rot}
=\hbar\omega_\mathrm{rot}\frac{\mathcal{J}_x}{\mathcal{J}_+^{(\mathrm{eff})}(\omega)}
=\hbar^2\frac{\langle J_x \rangle}{\mathcal{J}_+^{(\mathrm{eff})}(\omega)},
\label{energy}
\end{equation}
which is independent of $\omega_\mathrm{rot}$. 
Since the excitation energy of the first rotational state on the high-$K$ configuration 
in the rotor model is given by
\begin{equation}
E_{I=K+1}-E_{I=K}=\frac{\hbar^2}{\mathcal{J}}\left(K+1\right)
\label{energy_rot}
\end{equation}
derived from
\begin{equation}
E_I=\frac{\hbar^2}{2\mathcal{J}}\left(I(I+1)-K^2\right) ,
\label{rot}
\end{equation}
Eq.~(\ref{energy}) [$\langle J_x \rangle = K$ in the cases of 
$\gamma=-120^\circ$ or $60^\circ$] and Eq.~(\ref{energy_rot}) correspond to each 
other well for $K \gg 1$. 
In other words, $\mathcal{J}_+^{(\mathrm{eff})}(\omega)$ in our RPA formalism and 
$\mathcal{J}$ in the axially symmetric rotor model correspond to each other. 

 Marshalek gave an expression for multipole transition rates, which is valid for 
$I \gg 1$, in terms of the RPA wave function~\cite{ma2}. In the $M1$ case this reads 
\begin{gather}
B(M1:I\rightarrow I-1)=\frac{1}{2}\langle[i\mu_y+\mu_z,X_n^\dagger]\rangle^2 , \notag \\
\mu_{y(z)}=\sqrt{\frac{3}{4\pi}}\mu_N\left(g_l l_{y(z)}+g_s^{\mathrm{(eff)}} s_{y(z)}\right) ,
\label{m1}
\end{gather}
for the $n$-th phonon state. Hereafter we concentrate on the precession phonon. 
By equating this with the expression in the rotor model~\cite{bm}, 
\begin{equation}
B(M1:I=K+1\rightarrow K)
=\frac{3}{4\pi}\mu_N^2\left(g_K-g_R\right)^2K^2\langle I K 1 0|I-1 K\rangle^2 ,
\label{bm1_rot}
\end{equation}
we obtain the RPA $\left|g_K-g_R\right|$. Its sign is determined by that of the 
calculated $E2/M1$ mixing ratio. 

 Calculation is performed for all the high-$K$ (4, 6, 8 and 10QP) configurations that exhibit 
rotational bands; $K^\pi=13^-$, $14^+$, $15^+$, $18^-$, $21^-$, $22^-$, $25^+$, $28^-$, $29^+$, 
$30^+$ and $34^+$. The $K^\pi=13^-$, $14^+$ and $15^+$ are $2\nu2\pi$, $18^-$ is $2\nu4\pi$, 
$21^-$ and $22^-$ are $4\nu2\pi$, $25^+$, $28^-$ and $30^+$ are $4\nu4\pi$, and $29^+$ and $34^+$ 
are $6\nu4\pi$ 
configurations~\cite{w2,w3}. The model space is $N_{\mathrm{osc}} =$ 3 -- 7 for neutrons and 
2 -- 6 for protons. The strengths of the $\mathbf{l\cdot s}$ and $\mathbf{l}^2$ potentials are 
taken from Ref.~\citen{br}. The pairing gaps are assumed to be 0.5 MeV for 2QP and 0.01 MeV for 4 
and 6QP configurations both for neutrons and protons. 
The quadrupole deformation is chosen to be 
$\epsilon_2 =$ 0.235 that reproduces in a rough average the value $Q_0 =$ 7.0 eb that was assumed 
in the experimental analyses~\cite{w2,w3}. As for the spin $g$-factor, 
$g_s^{\mathrm{(eff)}}= 0.7 g_s^{\mathrm{(free)}}$ is adopted as usual. 
These mean that the choice of parameters in this 
work is semi-quantitative; we checked the robustness of the results with respect to their 
variations. In the cases symmetric about the $x$ axis considered here the results do not depend on 
$\omega_{\mathrm{rot}}$, while actual calculations are performed at 
$\hbar\omega_{\mathrm{rot}} = 0.001$ MeV. 

\begin{figure}
 \centerline{\includegraphics[width=7cm]{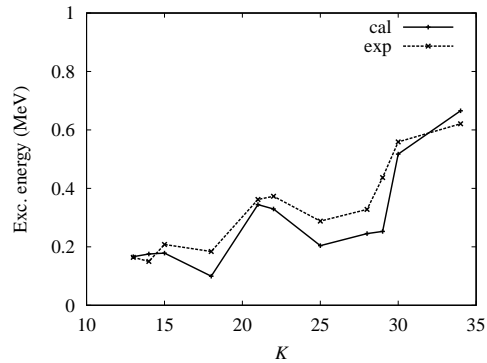}}
 \caption{Calculated and experimental excitation energies of the first rotational band members, 
          $E_{I=K+1}-E_{I=K}$. Data are taken from Refs.~\citen{w2,w3}. 
          }
 \label{fig1}
\end{figure}

 Figure~\ref{fig1} presents the calculated and observed relative excitation energies of the first 
rotational band members, $E_{I=K+1}-E_{I=K}$. 
Our RPA calculation reproduces their gross features well but with a close look one finds deviations 
at $K^\pi=18^-$, $25^+$, $28^-$, and $29^+$ that include the $\pi h_{9/2}$ orbital. Low calculated 
energies correspond to large 
moments of inertia [see Eq.~(\ref{energy})] and their largeness correlates with that of  calculated 
$Q_0$. The largeness of $Q_0$ indicates the shape polarization effect of this high-$j$ orbital to 
the prolate direction. 
As for the effect of the $\pi h_{9/2}$ orbital on the moment of inertia, see also 
Refs.~\citen{dra,fnsw}. 

\begin{figure}
 \centerline{\includegraphics[width=7cm]{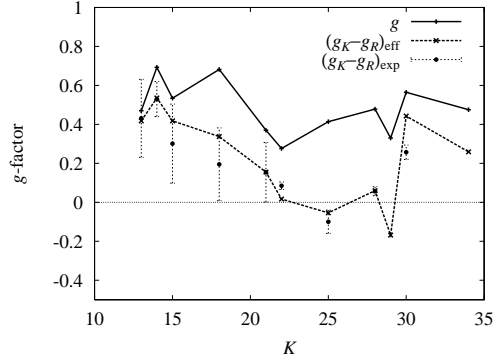}}
 \caption{Calculated intrinsic $g$ of high-$K$ configurations (solid curve) and 
          calculated (dashed curve) and experimental (points with error bars) $\left(g_K-g_R\right)$. 
          Data are taken from Refs.~\citen{w2,w3}. 
          }
 \label{fig2}
\end{figure}%

\begin{figure}
 \centerline{\includegraphics[width=7cm]{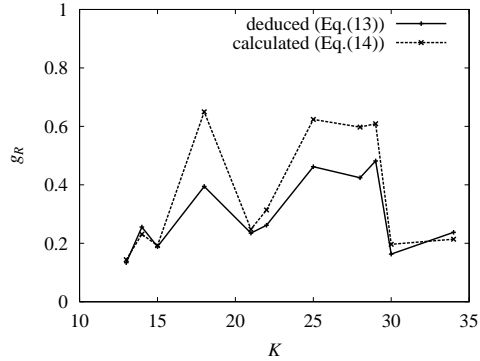}}
 \caption{$g_R$ of high-$K$ rotors deduced using Eq.~(\ref{grel}) (solid curve) and calculated 
          using Eq.~(\ref{gr_j}) (dashed curve). 
          }
 \label{fig3}
\end{figure}

 In Fig.\ref{fig2} we compare the RPA $\left(g_K-g_R\right)$ extracted from 
the calculated $B(M1; I=K+1 \rightarrow K)$ and the observed one extracted assuming $Q_0 =$ 7.0 eb 
from the branching ratios of available lowest transitions in respective rotational bands. 
Their agreement is semi-quantitative. In this figure the calculated intrinsic 
$g = \sqrt{\frac{4\pi}{3}}\langle \mu_x \rangle/(\langle J_x \rangle\mu_N)$ of high-$K$ configurations are also shown. 
The two calculated curves roughly correlate. 
According to the relation~\cite{bm}
\begin{equation}
g_R=g-(g_K-g_R)\frac{K^2}{I(I+1)} 
\label{grel}
\end{equation}
with $I=K+1$, $g$ almost coincides with $g_K$. Consequently, 
the difference between the two curves essentially corresponds to the effective 
$g_R$ for high-$K$ cases. 
Thus, this correlation suggests a possibility to deduce the effective $g_R$ of 
the considered high-$K$ configurations by substituting the RPA $\left(g_K-g_R\right)$ to 
Eq.~(\ref{grel}).  
Its average value is about 0.29 as seen from Fig.\ref{fig3}. 
This value may be a rough measure of a property of high-$K$ rotors. Moreover, an interesting 
feature is that there are considerable variations and
those for the configurations including the $\pi h_{9/2}$ orbital are larger than others.
In order to see it more closely, in Fig.\ref{fig3} we compare them with those calculated from 
an approximate relation~\cite{ba}
\begin{equation}
g_R=\frac{\mathcal{J}_\pi}{\mathcal{J}_\nu+\mathcal{J}_\pi} ,
\label{gr_j}
\end{equation}
where the neutron and proton part of the effective inertia, 
the upper sign of Eq.~(\ref{effiner}), 
are substituted to $\mathcal{J}_\nu$ and $\mathcal{J}_\pi$. 
It is clear that the $g_R$ values deduced from Eqs.~(\ref{grel}) and (\ref{gr_j})
correspond to each other very well, although those from Eq.~(\ref{gr_j}) are much larger
for the configurations in which the $\pi h_{9/2}$ orbital is occupied:
The contribution to the moment of inertia from the $\pi h_{9/2}$ orbital
is large and overestimated in the calculation as mentioned before in the case
of excitation energies.

 Finally we compare the deduced $g_R$ of high-$K$ rotors above and that of the ground 
state band. We calculate $g$ of the $2^+$, $4^+$, and $6^+$ states, which to the zeroth 
approximation plays a role of $g_R$ for the nearby $K \neq 0$ configurations, at 
$\hbar\omega_{\mathrm{rot}} =$ 0.053, 
0.119, and 0.176 MeV, respectively, with $\epsilon_2 = 0.235$, $\gamma = 0$, 
and the odd-even mass differences $\mit\Delta_n =$ 0.883 MeV and $\mit\Delta_p =$ 1.026 MeV 
as pairing gaps. The results are $g =$ 0.218, 0.216, and 0.214, respectively, 
which almost coincides with the average for the configurations that do not include the $\pi h_{9/2}$ 
orbital. This indicates that high-$K$ and low-$K$ rotors are similar unless the shape driving 
$\pi h_{9/2}$ orbital is included. 

 To summarize, we have numerically verified that the random phase approximation performed 
on high-$K$ multi-quasiparticle configurations leads to a rotor picture, as previously 
discussed via $E2$ properties by Andersson \textit{et al.}~\cite{ander}, by calculating excitation 
energies and $M1$ properties. Next we have deduced the effective $g_R$ of the high-$K$ rotors 
and compared them with those of the low-$K$ rotor near 
the ground state. A more detailed investigation is under progress.

\end{document}